%% file: main.tex
\documentclass[format=sigplan, 10pt]{acmart}
\usepackage{url}
\usepackage{color}
\usepackage{float}
\usepackage{graphicx}
\usepackage{multirow}
\usepackage{array}
\usepackage{amsmath}
\usepackage{multicol}
\newcommand{\round}[1]{\ensuremath{\lfloor#1\rceil}}
\usepackage{comment}
\usepackage[normalem]{ulem}
\usepackage{tikz}
\usepackage{enumerate}
\usepackage{multirow}
\usepackage{array}
\usepackage{listings}
\usepackage[ruled,vlined]{algorithm2e}
\usepackage{amsmath}
\usepackage{amsfonts}
\usepackage{pifont}
\usepackage[ragged2e]{}
\usepackage{hyperref}
\usepackage{breakurl}
\usepackage{float}
\usepackage[utf8]{inputenc}
\definecolor{gitadd}{HTML}{00A64F}
\definecolor{gitdel}{HTML}{c94238}

\input{myfunctions}

\begin{document}

\renewcommand\footnotetextcopyrightpermission[1]{}


\title{Programming and Deployment of Autonomous Swarms using Multi-Agent Reinforcement Learning}

\author{Jayson Boubin}
\email{boubin.2@osu.edu}
\affiliation{
    \institution{Ohio State University}
    \city{Columbus}
    \state{Ohio}
    \country{USA}
}

\author{Codi Burley}
\email{burley.66@osu.edu}
\affiliation{
    \institution{Ohio State University}
    \city{Columbus}
    \state{Ohio}
    \country{USA}
}

\author{Peida Han}
\email{han.1242@osu.edu}
\affiliation{
    \institution{Ohio State University}
    \city{Columbus}
    \state{Ohio}
    \country{USA}
}

\author{Bowen Li}
\email{li.7652@osu.edu}
\affiliation{
    \institution{Ohio State University}
    \city{Columbus}
    \state{Ohio}
    \country{USA}
}

\author{Barry Porter}
\email{b.f.porter@lancaster.ac.uk}
\affiliation{
    \institution{Lancaster University}
    \city{Lancaster}
    \country{United Kingdom}
}

\author{Christopher Stewart}
\email{cstewart@cse.ohio-state.edu}
\affiliation{
    \institution{Ohio State University}
    \city{Columbus}
    \state{Ohio}
    \country{USA}
}

\renewcommand{\shortauthors}{Boubin et al.}

\thispagestyle{firstpage}
\pagestyle{plain}

\input{abstract}

\settopmatter{printacmref=false}

\maketitle
%

\input{introduction}

\input{Design}

\input{AppSoftware}

\input{OnlineLearning}
\input{Implementation}
\input{Results}
\input{Discussion}
\input{relatedwork}
\input{conclusion}

{\small
\bibliographystyle{abbrv}
\bibliography{references}
}

\end{document}

%% file: myfunctions.tex
\usepackage{upgreek}

\newcommand{\MYcomment}[1]{}

\newcommand{\MYnote}[1]{}

\newcommand{\MYlt}{\textless}

\newcommand{\MYfigureref}[1]{Figure~\ref{#1}}

\newcounter{MYtablecntr}
\addtocounter{MYtablecntr}{1}

\newcommand{\MYlabel}{\small {$\bullet$}}

\newcounter{MYenumctrtwo}
\newenvironment{MYenumwide}{\begin{list}{\arabic{MYenumctrtwo}.}{%
\usecounter{MYenumctrtwo}%
\setlength{\topsep}{2pt plus 0pt minus 0pt}%
\setlength{\itemsep}{2pt plus 0pt minus 0pt}%
\setlength{\parsep}{2pt plus 0pt minus 0pt}%
\setlength{\parskip}{0pt plus 0pt minus 0pt}%
\setlength{\parindent}{0pt }
\setlength{\leftmargin}{15pt}%
}}{\end{list}}

\newcounter{MYenumctr}

%% file: abstract.tex
\begin{abstract}


Autonomous systems (AS) carry out complex missions by 
continuously observing the state of their surroundings 
and taking actions toward a goal. Swarms of AS working 
together can complete missions faster and more 
effectively than single AS alone. To build swarms today, 
developers handcraft their own software for storing, 
aggregating, and learning from observations. 
We present the Fleet Computer, a platform for 
developing and managing swarms. 
The Fleet Computer provides a programming paradigm that
simplifies multi-agent reinforcement learning (MARL) -- an emerging class of algorithms that coordinate swarms of agents. 
Using just two programmer-provided functions \textit{Map()} and \textit{Eval()}, 
the Fleet Computer compiles and deploys swarms and 
continuously updates the reinforcement learning models 
that govern actions.  To conserve compute resources,
the Fleet Computer gives priority scheduling to models
that contribute to effective actions,  drawing a novel
link between online learning and resource management.
We developed swarms for unmanned 
aerial vehicles (UAV) in agriculture and for video analytics on urban traffic. 
Compared to individual AS, our swarms achieved speedup of 
4.4X using 4 UAV and 62X using 130 video cameras. 
Compared to a competing approach for building swarms that is widely used in practice,
our swarms were 3X more effective, using 3.9X less energy.

\end{abstract}

%% file: introduction.tex
\vspace{-0.1in}
\section{Introduction}

Autonomous systems (AS) continuously sense their 
surroundings, model their current state and take 
actions toward a goal.  Edge computing, a paradigm 
where compute resources are provisioned near sensing devices, 
has propelled AS in a wide range of industries~\cite{mohan2020pruning,boubin2019autonomic,hu2020hivemind,zhou2020uav}. 

Groups of AS working toward a common goal 
are called {\em swarms}.  Compared to AS working 
alone, swarms can speed up mission execution.
First, missions can be partitioned into tasks 
that swarm members execute in parallel.  Second, swarm 
members can share observations of their 
surroundings to help other members take 
effective actions. Today, human programmers
manually partition missions for swarm operations
and each swarm member uses pre-programmed behaviors.
For example, in precision agriculture, such 
{\em automated swarms} divide a crop field among 
multiple UAV and each UAV executes pre-programmed
scouting routines on its region~\cite{barrientos2011aerial}. 
Recent research provides automated partitioning
and fault tolerance for automated swarms~\cite{hu2020hivemind}.
However, pre-programmed behaviors fundamentally waste 
resources by collecting and processing data of low 
value relative to the mission.  Further, by 
limiting autonomy, data sharing
between members can not improve efficacy.
 
Multi-Agent Reinforcement Learning (MARL) is an 
emerging class of reinforcement learning algorithms 
where agents cooperate to maximize a reward.  
Agents learn their own reinforcement-learning 
policies, but they can also learn from the actions 
and outcomes of other agents.  MARL algorithms
applied to AS swarms can speedup missions via
partitioning (like the automated approach above) and 
via efficacy (i.e., taking better actions).  Further, recent 
research on MARL algorithms provides
provable guarantees and strong empirical results~\cite{lin2020collaborative, busoniu2008comprehensive, zhang2018networked,cui2019multi}.
However, MARL systems are not simple to develop and manage;
they require infrastructure for AS workflows, selecting MARL algorithms and reward functions, and
data management policies.  Developers must create this infrastructure by hand and incorporate it into a real-world system. 
The result is that, despite their potential, these algorithms rarely go beyond theoretical studies or highly specialised applications with bespoke components.

We present the \textit{Fleet Computer}, a platform 
for developing and managing MARL-driven swarms. 
To program AS in the Fleet Computer, developers provide
two functions: {\em Map()} and {\em Eval()}.  Map() converts
sensed observations (e.g., images) to application-specific feature vectors.
Eval() evaluates system performance towards autonomy goals.
In addition, developers also provide hardware resources 
and mission configuration settings, e.g., allowed actions and goals.  
With these inputs, the Fleet Computer compiles MARL models that govern autonomous actions.
Then, during execution, the Fleet Computer aggregates 
data from swarm members and retrains the models governing actions.
However, retraining stresses limited computational resources at edge sites.
The Fleet Computer prioritizes retraining for models most
useful for effective actions, making a novel link between
online learning and resource management.
The Fleet Computer also expands and contracts edge 
compute resources via duty cycling to save energy and 
reducing over provisioning.  

Just as MapReduce~\cite{dean2008mapreduce} simplified parallel data processing, 
the Fleet Computer demystifies fully autonomous swarms.  In addition
to its novel programming paradigm, the Fleet Computer provides end-to-end control, deployment, and scheduling of an entire swarm-support infrastructure -- from individual swarm units, such as a UAV, to the supporting heterogeneous compute
resources at the edge. 
 
 
The Fleet Computer comprises 3 main contributions:
\begin{MYenumwide}
\item A programming model and toolchain that allows users to easily build MARL-driven swarms. 
\item A novel, online-learning approach to aggregate observations from swarm members 
      and dynamically update models governing actions.
\item A runtime platform that manages, schedules and duty-cycles edge compute resources, simplifying deployment. 
\end{MYenumwide}

We used the Fleet Computer to build both aerial crop scouting and video analytics systems, demonstrating its broad applicability to significantly different problems. We developed a swarm of unmanned aerial vehicles used to predict crop yield from flyover images. We also developed a taxicab tracking workload using a swarm of autonomous cameras. By providing a simple way to take advantage of state-of-the-art MARL algorithms, our crop yield mapping application
outperforms state of the art yield mapping, improving mapping times by 4.4X using 3.9X less UAV power. Similarly, our vehicle tracking workload built using the Fleet Computer outperforms prior work, tracking taxis up to 62X faster as a swarm compared to centralized processing while maintaining good performance over time by updating models regularly. Using these applications, we demonstrate that
a compute cluster running Fleet Computer software can easily scale up from a single UAV, camera, or other agent to a complete swarm that can learn from its actions while dynamically allocating resources to fit
demand and save precious power at the edge. 


The remainder of the paper is organized as follows:
Section~\ref{sect:background} describes at a high level the overarching design of the Fleet Computer. 
Section~\ref{sect:design} details our programming model, which allows users to easily build autonomous swarms that perform well.
Section~\ref{sect:online} discusses the Fleet Computer's runtime, which includes a priority-based online learning approach to maintain model performance as environments change, and cluster autoscaling which saves edge power without sacrificing performance. 
Section~\ref{sect:impl} covers two Fleet Computer implementations: video analytics and autonomous crop scouting. 
Section~\ref{sect:results} presents results for the Fleet Computer on both applications. 
Section~\ref{sect:rw} presents related work and Section~\ref{sect:conclusion} provides conclusions.

%% file: Design.tex
\section{Background}
\label{sect:background}
Self-driving cars~\cite{lin2018architectural, urmson2008autonomous}, 
UAV~\cite{boubin2019managing,sanchez2016aerostack,zhang2012application},  
and other autonomous systems (AS)~\cite{fahimi2009autonomous,siegwart2011introduction,pfeiffer2017perception,bouman2020autonomous} 
are transforming society.
Investments in self-driving cars exceed \$56B~\cite{uavmarket2020}. 
UAV are transforming delivery, infrastructure monitoring, and surveillance~\cite{primeair,boubin2019managing,al2017vbii,semsch2009autonomous, hu2020hivemind}.
Autonomous cameras with mobile gimbals are powering traffic monitoring and smart cities~\cite{jain2020spatula}. 
In this paper, the term {\em autonomous} describes systems 
that sense their surroundings, infer their state,
and take actions toward mission goals {\em without human intervention.}

By design, AS execute in unfamiliar surroundings.  
Their efficacy depends on how well their actions align to mission goals.
Broadly, AS can be characterized by two key design choices:
(1) Do they learn from their own observations? 
And (2) do they learn from the observations of other AS?
For many AS used in practice today, the answer to both
questions is no.  These {\em pre-programmed AS} repeat
the same routine across all missions, often taking 
unneeded and wasteful actions.

With reinforcement learning, AS can learn from 
prior observations~\cite{boubin2019managing,lin2018architectural,kato2015open,faessler2016autonomous,sanchez2016aerostack}. 
However, often in practice, AS observe the world too
slowly to capture enough data for learning, especially
in non-stationary contexts where the best actions change.
Multi-agent reinforcement learning (MARL) addresses this problem
by aggregating data from multiple AS (i.e., swarms).
Previously, the challenge for MARL-driven AS has 
been deciding from which swarm members to learn, but
recent research has developed online algorithms to 
explore data aggregations that have provable guarantees and/or strong empirical
results~\cite{lin2020collaborative, busoniu2008comprehensive, zhang2018networked,cui2019multi}. 
However, even with recent research, swarms require
additional compute resources compared to single-agent AS.  
To share data, swarms members must be networked, e.g., via 
hubs~\cite{vasisht2017farmbeats,bastug2017toward} or wireless~\cite{hassan2019edge,wan2019toward}. 
Exploring data aggregations also requires additional compute resources.

\vspace{0.05in}
\noindent \textbf{Autonomous UAV for Crop Yield Modeling:} 
To illustrate these concepts, consider a farmer that owns 4 UAV 
controlled via apps running on 4 tablets.  The tablets and an 
edge-hub desktop share a wireless Internet connection.
The mission is to map the expected crop yield for 
each $0.01\times$ acre lot in a $1,000$ acre field with accuracy above $80\%$.
For the approach most commonly used in practice, 
the farmer would use pre-programmed routines
to exhaustively capture images from $100,000$ waypoints.
This approach is slow and unnecessarily exceeds accuracy goals.
A pre-programmed swarm could split the mission into 4 parallel tasks.
With reinforcement learning, AS can visit significantly
fewer waypoints~\cite{boubin2019autonomic,zhang2020,sanchez2016aerostack}
by continuously modeling the expected accuracy of their map and 
prioritizing valuable waypoints to visit next.
MARL-driven AS improve upon naive reinforcement learning by using 
shared images to visit fewer waypoints.
However, the computational load for reinforcement learning and MARL
exceeds the capacity of UAV processors and tablets, requiring resources
of an edge hub or the cloud.

%% file: AppSoftware.tex

%




\section{Design}
\label{sect:design}

\begin{figure}[t]
 \centering
  \includegraphics[scale=0.90]{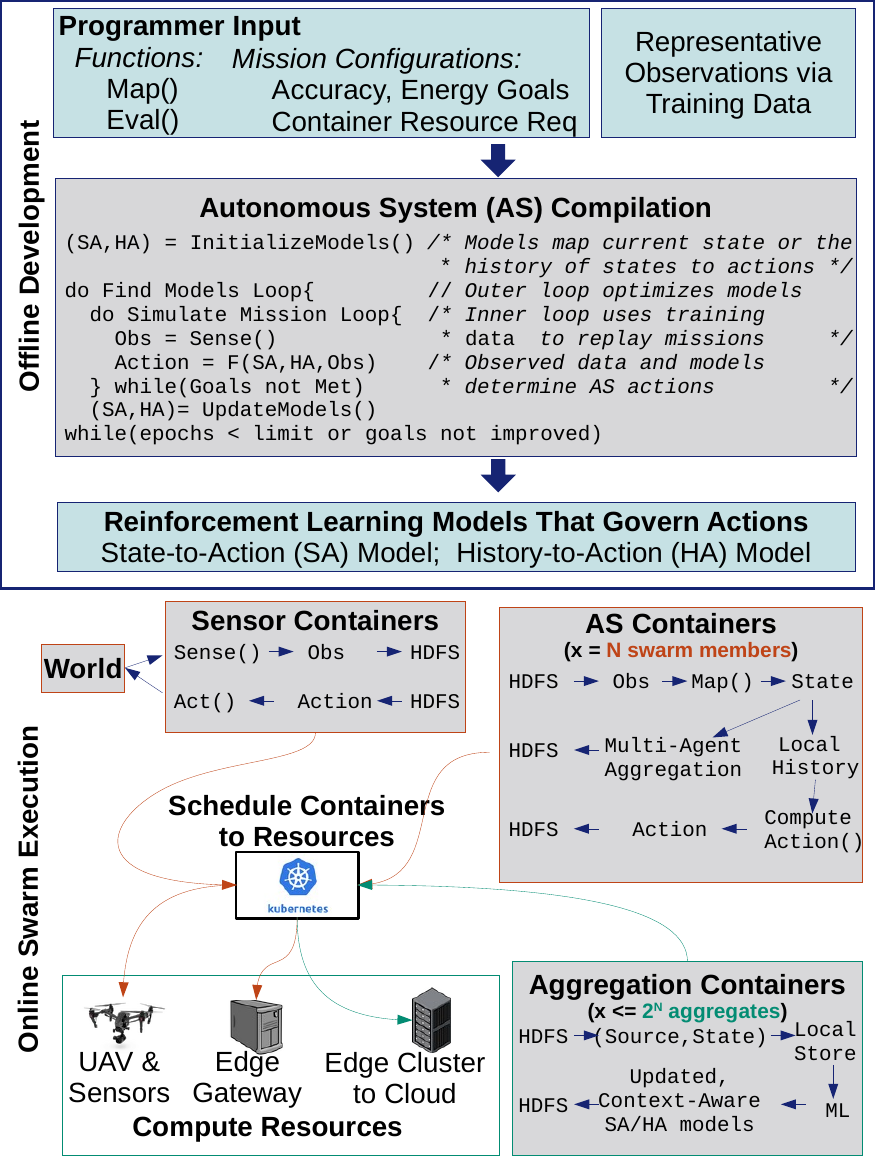}
 \caption{The architecture of the Fleet Computer.}
  \label{fig:arch}
\end{figure}

As shown in Figure~\ref{fig:arch}, the Fleet Computer
is an end-to-end platform for autonomous swarms, covering
the development of AS, their workflow and coordination, and
their execution on edge computing devices.

To create an AS, developers implement two functions, 
Map() and Eval(), and specify mission configuration.
Map() functions convert quantized input from sensing devices
(e.g., cameras, GPS, etc.) into a feature vector 
that represents the state of the AS. Eval() functions
aggregate all outputs from Map() invocations during an
epoch and assess the extent to which the mission has been completed.
The mission configuration defines key parameters that
developers can adjust across swarm applications.
Figure~\ref{fig:arch} depicts three examples of mission 
configuration settings.
First, developers can provide thresholds to determine
when missions are complete. By default, the Fleet 
Computer supports thresholds on accuracy and energy 
usage. Second, developers
specify the type of resources needed for sensing, 
processing, and data aggregation by stipulating quantitative requirements, 
such as CPU and memory required for Map() execution.  Third, users provide 
qualitative requirements, such as support for specific 
Action Drivers. Action Drivers support a cyber-physical system's actions, e.g., take off, 
land, sense, and fly to waypoint. 

The Fleet Computer compiles these inputs to create
AS workflows for sensing surroundings and taking actions.
Here, the challenge is to decide which actions to take 
after running Map() and Eval().  
The Fleet Computer automatically builds state-to-action (SA)
models and history-to-action (HA) models by (1) replaying 
data from prior execution contexts, and
(2) learning effective actions that improve Eval() outcomes.
SA models convert a single map output
to actions whereas HA models convert multiple observed outputs.

The Fleet Computer models MARL-driven AS as three 
asynchronous components: Sensors, AS Workflow, and Data Aggregation.
These components execute in shared-nothing containers 
connected via distributed storage (e.g., HDFS~\cite{shvachko2010hadoop}).
Containers are replicated to support swarms.  Precisely, 
a swarm of size $N$ will comprise $N$ Sensor and AS containers
and up to $2^{N}$ Data Aggregation containers 
(representing every possible combination of aggregations).
Clearly, Data Aggregation containers impose computational
demands that exceed system capacity--- a challenge common
to all MARL-driven approaches~\cite{lin2020collaborative, busoniu2008comprehensive, zhang2018networked}.  
For small swarms ($N\leq8$),
The Fleet Computer deploys all Data Aggregation containers
and employs a novel, priority-based online learning and 
scheduling to manage compute demand.  For larger swarms,
the Fleet Computer supports developer-provided heuristics to limit
aggregation.

In this section, we first provide a 
rigorous primer on MARL algorithms.  Then, we introduce
the specification of Fleet Computer applications, i.e., swarms of AS.  
Finally, we describe each of the key functions and models listed
above.

\subsection{MARL Primer}

Broadly, reinforcement learning approaches determine 
a policy $\pi^*$ that approximates an optimal solution 
to a Markov Decision Process (MDP).  
MDPs comprise States $S$, Actions $A$, 
a transition probability function $P \rightarrow S x A \rightarrow \Delta(S)$, 
a reward function $R(s_i,a_i,s_{i+1})$ defining the immediate reward an agent receives from performing an action, 
and a discount factor $\gamma$~\cite{zhang2019multi}. 
A policy $\pi$ is a means of determining which action to take in a given state to transition to another state. The optimal policy $\pi^*$ is the policy that results in the set of state transitions that maximizes overall reward.  
\begin{equation}
    MDP: (S, A, P, R, \gamma)
\end{equation}
MDPs are a useful tool for solving simple state transition problems, but fleet missions are too complex for this framework. Consider the crop yield modeling example from Section 2. Crop scouting is a complex workload where the execution context changes over time and reward is difficult to define. If the system is rewarded based on the estimated yield it predicts in each state, it will likely over or under-predict yield based on how reward is assigned. Furthermore, even if reward is properly assigned, it is likely that states and transition probabilities will change over time as crops grow and conditions change. In many cases, an optimal policy is nearly impossible to determine a priori. For these tasks, reinforcement learning is used to develop ${\pi^*}' \approx \pi^*$, an approximately optimal policy for navigating execution contexts while maximizing reward.

There are many methods for approximating ${\pi^*}$ through reinforcement learning including value based methods like Q-learning, policy based methods like actor-critic RL, and analogous deep methods like Deep Q-Networks and Deep Deterministic Policy Gradients~\cite{watkins1992q,grondman2012survey,mnih2015human,lillicrap2015continuous}. Throughout this paper, we will use Q-learning as a basis for MARL, but other techniques fit into our programming model as well.
Q-learning is a value-based method for reinforcement learning which uses a Q-function to determine ${\pi^*}'$. $Q(S_i,A_i)$ is the Q-function which predicts the expected value (Q-value) of an action $A_i$ taken at state $S_i$. In practice, Q-values are stored in a Q-table $Q[S,A]$ indexed by state-action pairs. When an action is performed, the Q-table is updated using the bellman equation shown in equation 2, which uses dynamic programming to update the Q-value for a state-action pair based on the reward for a given action, plus expected reward of all future actions modified by learning rate $\alpha$ and discount factor $\gamma$. Properly informed Q-tables and other RL mechanisms solve MDPs with high accuracy by learning ${\pi^*}'$, allowing them to learn complex behaviors through exploration.

\begin{equation}
\begin{aligned}
    Q(s_i,a_i) = (1-\alpha)*Q(s_i,a_i) + \\ \alpha[R(s_i,a_i,s_{i+1}) * \gamma max(Q(s_{i+1},a_{i+1}))]
\end{aligned}
\end{equation}

This process translates quite well to multi-agent systems. Transitioning a reinforcement learning algorithm to a multi-agent domain can involve constructing careful global reward functions~\cite{chang2004all}. Another approach, team-average reward~\cite{kar2013cal,doan2019finite}, maximizes the reward received by the system given agents with different and potentially discordant reward functions. MARL algorithms of this type are called Markov Games (MGs)~\cite{littman1994markov}. MGs, shown in equation 3, expand the MDP by adding multiple agents, defined by $N\geq1$. Each agent $i \in N$ has its own action set $A_i$ and reward function $R_i$. Similar to MDPs, the solution to the MG is policy $\pi^*$, the set of state transitions that maximizes reward. Much work has also been done with networked agents~\cite{zhang2018fully, qu2020scalable, zhang2018networked}, agents within a MG that communicate over some time-varying network, may have individual reward functions, and may require data privacy.

\begin{equation}
MG = (N, S, A^i_{i \in N}, P, R^i_{i \in N}, \gamma)
\end{equation}

Given this specification, the Fleet Computer should accommodate different MARL algorithms using the same base components while assuring that these algorithms fit within the fleet framework. To allow the design and deployment of MARL algorithms for real-world AS, the Fleet Computer takes some of these base MARL components as inputs and generates others offline.

\subsection{The FleetSpec}

\begin{equation}
    Fleet Spec: (N, S, A^i_{i \in N}, Map(), Eval(),\mathbb{C})
    \label{eq:fleetspec}
\end{equation}

Equation~\ref{eq:fleetspec} presents the minimum specification 
for Fleet Computer applications, called the FleetSpec.
Similar to a Markov Game, the FleetSpec accepts a number of 
agents $N \geq 1$, states $S$, and action sets $A_i$ for each agent. 
States are non-injective and surjective mappings from action 
sequences to integers $<a^t_{t=0},a^t_{t=1}...a^t_{t=T}> \rightarrow \mathbb{Z}$ where $a^t \in A^i_{i \in N}$.
States affect the behavior of action drivers.  For example,
if a vehicle is at the eastern edge of allowed states, the
command {\em go East} is muted by the action driver.  
Unlike Markov Games, FleetSpec eschews state transition probability
and reward functions. First, Fleet Computer developers can
reuse action drivers created by others.  The action drivers 
may support states about which the developers are unaware, making
state transition models incomplete.  
Second, constructing mathematical reward functions is challenging. 
Real-world AS take on missions that involve complex,
domain-specific knowledge.  The value of their actions 
can be subtle and may depend on prior actions, requiring
complex non-linear reward functions that overly complicate 
the development of AS.

Instead, the Fleet Computer learns state-to-action ($SA$) models 
and history-to-action models ($HA$) automatically through training and
reward shaping.  These models represent an approximation of ${\pi^*}'$.  The Map() and Eval() functions, coupled with
representative observations from prior missions $\mathbb{C}$, suffice
to compile initial MARL models and reward functions.
The remainder of this section details the compilation process.

\subsection{Map() and Eval() Functions}
Like in MapReduce, Map() functions in the Fleet Computer structure input data.  
AS get their input data from 
sensor containers.  The output is a feature vector, called a state-space vector (SSV),
that describes sensed observations in the system's 
current state. Precisely, let $D_j$ be the sensor data
observed in state $S_i$, $Map(D_j)$ directly translates 
observed sensor data to a SSV, as shown below.
\begin{equation}
    D = <d_1, d_2, ... d_n> 
\end{equation}
\begin{equation}
     Map(D) = SSV = <f_1, f_2, ... f_m>
\end{equation}
By emitting a structured SSV, Map() functions in the Fleet Computer can
compose multiple {\em extractor functions} that process a portion
of the sensed data $D' \subset D$ and emit part of the SSV.
Extractor functions are shown in \MYfigureref{fig:map-func} for our crop scouting example. $Map(D_j)$ for crop scouting provides data $D_j$ to extractors including $ExG()$ which determines excess green~\cite{khanal2018integrating} (a metric for predicting crop yield), $LAI()$ which estimates the leaf area index~\cite{raj2021leaf} of crops in the image, and a CNN which counts the number of corn stalks in the image among other extractors. Each of these extractors provides important information about the execution context that can be used to both build final yield maps and predict optimal actions for sampling. Each of these extractors return one or more floating point values which are added to the final SSV. 

\begin{figure}[t]
    \centering
    \begin{lstlisting}
 func[] extractors = [ExG(), LAI(), 
                CornCountCNN(), ...]
 float[] Map(Obj data) {
   float[] SSV = [];
   for(e in extractors) {
     SSV.append(e(data));
   }
   return SSV;
 }
 
 Obj[] Eval(float[][] FS, float[][] perf) {
   finished = checkGroups(FS, HA)
   map_final = []
   if(finished) {
     map_gt = buildMap(FS)
     map_final = extrap(map_gt)
   }
   P = buildMetrics(perf)
   return [finished, P, map_final]
 }
 
    \end{lstlisting}
    \caption{Map and Eval function pseudocode for crop scouting.  
    }
    \label{fig:map-func}
\end{figure}

Eval() determines whether and to what degree an AS has accomplished its goal. For some AS, this can be as simple as reaching a certain state. For others, like the autonomous crop scouting example, goal evaluation is more difficult. Depending on the size and type of field being modeled, the goal may be to make the most accurate yield map possible within some timeframe, cost, or sampling coverage. 

\begin{equation}
    FS = \{SSV_1, SSV_2, SSV_3 ... SSV_n\}
\end{equation}
\begin{equation}
    P = \{\rho_1, \rho_2, \rho_3 ... \rho_m\}
\end{equation}
\begin{equation}
   Eval(FS, m) = P
\end{equation}

Eval(), defined above, accepts a feature space $FS$ comprised of $n \geq 1$ 
SSVs and a set of system-level metrics $m$, and outputs an evaluation $P$ which 
includes $x \geq 1$ evaluation metrics $\rho_1 .. \rho_m$. 
The number of state-space vectors and evaluation metrics required is task and goal dependant.

For our crop scouting UAV example, many or all sensed SSVs from a mission may be needed to build an accurate yield map. Evaluation metrics for crop scouting may include system execution time, accuracy, energy expenditure, monetary cost and any other metric that determines real-world performance. Eval() pseudocode for crop scouting is shown in ~\MYfigureref{fig:map-func}.

Eval accepts two parameters, a featureset of all SSVs collected by the system, and a set of performance metrics for each agent. First, we use the HA model to determine whether the mission has concluded by exploring all state groups, a process detailed in section 3.4. If the mission is finished, a final yield map is generated by converting FS into a ground-truth yield map by mapping ExG from each SSV into a yield prediction and mapping it onto the execution context. Then, unvisited zones are extrapolated using an approach from prior work~\cite{zhang2020}. Next, Eval() uses performance metrics from every agent to build a set of global metrics that can be used for goal evaluation, $P$. Finally, Eval() returns a Boolean stopping condition based on HA, evaluation metrics, and potentially a final yield map which will be returned to the user.

\subsection{State-To-Action and History-To-Action Models}
Designing MARL policies and reward functions is a complicated problem with many situational solutions. Each action taken by an agent must be assessed by a policy before it is selected, and assigned reward if it is taken. The Fleet Computer's model training and reward shaping steps simplify this process, allowing users without predefined models or reward functions the option to build them automatically.

Training reinforcement learning policies generally requires releasing an agent into an execution context and allowing it to explore, building a policy using a reinforcement learning model from the reward it garners from its actions. Characterizing entire real-world execution contexts for training models is a well-explored problem. Point-cloud datasets have long been used in robotics and computer vision for simulating navigation and robotic manipulation in real-world environments~\cite{OaklandDS, NYUDS, sydney, SUM}. Data sets for self-driving vehicles~\cite{KITTI, DeepDrive, Level5} and video analytics~\cite{PORTO, BEIJING} provide labeled video streams of fully explored environments used to train algorithms in both domains. A similar approach, Autonomy Cubes, provides spatially and temporally linked images in the form of hypercubes representing completely explored execution contexts regularly used to develop autonomous UAV workloads~\cite{boubin2019managing, yang2020adaptive,zhang2020}. 
\begin{table}
\centering
\footnotesize
\begin{tabular}{ |p{1.4cm}||p{1.2cm}|p{1.1cm}|p{1.35cm}|p{1.4cm}| }
 \hline
 {\bf Name} & {\bf Linkage} & {\bf Data Type} & {\bf Devices} & {\bf Use-Cases} \\
 \hline
 Point clouds & Spatial & RGBD, ASCII & Robots & Indoor and outdoor Navigation \\
 \hline
 Video streams & Spatial, Temporal & Video & Self-driving cars, Video analytics & Trans-portation, Tracking,\\
 \hline 
 Autonomy cubes & Spatial, Temporal & Images, Video & UAV & Precision agriculture, Rescue, Photography\\
 \hline
\end{tabular}
\caption{\label{tab:terms} Execution context data sets and their use cases.}
\end{table}

Any of these execution context representations can be used to train MARL models for the Fleet Computer. In the FleetSpec, these execution contexts are defined as $\mathbb{C}$. Map() and Eval() combined with autonomy goals allow agents to navigate these environments just like the real world, with Map() extracting data from a given state, and Eval() determining whether autonomy goals are sufficiently accomplished. These functions do not, however, constitute a policy. The policy for navigating these environments is defined by State-to-Action and History-to-Action models detailed below. 

A State-to-Action model (SA) is a blank or pretrained MARL model $SA \approx \pi^*$. SA accepts a $SSV$ as input and returns an action. SA is not strictly tied to any RL or MARL model format, and is meant to be general. By default, the Fleet Computer supports Q-learning, but other model types, like DQNs or DDPGs can be substituted as an SA baseline. SA can also be provided pre-trained, or the Fleet Computer can instantiate it from a blank model and train it with data from $\mathbb{C}$. The only restriction placed on SA by the Fleet Computer is that it must accept an $SSV$ produced by Map() as an argument.

History-to-Action() models provide spatial support to SA models. An HA model determines if an AS has accomplished its goal within its execution context. As an AS explores its execution context, some states of low relevance can be excluded or extrapolated to conserve time, energy, and compute resources. MARL models generally rely solely on reinforcement learning to determine which states to search, but recent video analytics work~\cite{jain2020spatula} demonstrates how spatial and temporal correlations can help prune search spaces effectively. In response to this work, we propose that groups of states (state groups) be allocated in conjunction with HA models to better search across execution contexts. 

A state group is a collection of states within the execution context, with each state belonging to one or more state groups. State groups provide spatial bounds on AS. Often, phenomena sought by AS is spatially correlated~\cite{zhang2020, jain2020spatula}. By defining spatial bounds, AS can limit their searches, saving time and resources. HA determines the number of states explored in each group. After a state in the state group is visited, the feature vectors of each visited state in that group are provided to HA to determine whether to explore another state in that group, or to visit another group. If the system decides to visit another state group, all unexplored zones in the group are either disregarded or predicted by Eval(). HA is defined as follows. 

~\begin{equation}
    HA(FS) = 
    \begin{cases}
        U > T_u \: \text{or} \:  V > T_v  & 0 \\
        else & 1
    \end{cases}
\end{equation}
\begin{equation}
    U = \sum_{i=0}^{n} R(S_i,A_i,S_{i+1})
\end{equation}
\begin{equation}
    V = \|FS\|
\end{equation}

$HA$ accepts a feature space $FS$ comprised of one or more state-space vectors $SSV_1..SSV_n$ from the same state group. $HA(FS)$ returns a Boolean value representing whether to continue exploring a state group (0) or to move on and explore another (1). This decision is made using $U$, the total utility of the feature space, and $V$, the size of the feature space (number of visited states in the group). The utility of a feature space is the aggregate reward received from all state transitions within that space. If $U$ is above $T_u$, the utility threshold, or $V$ is above $T_v$ the visited states threshold, all remaining states in the group are ignored or predicted by Eval() and another state group is visited. The two thresholds strike a balance between allowing the MARL algorithm time to find high reward state transitions and limiting the total number of states visited to maintain efficiency. 

These thresholds, as well as the means for determining reward, are difficult to determine a priori. Rigorous construction of high-quality reward functions is still an open problem, and could be difficult for users attempting to apply autonomy to a new domain or in an unclear application~\cite{kilinc2018multi, zhang2019multi, zou2019reward}. Work has been done on reward shaping~\cite{zou2019reward, lin2018multi, grzes2017reward}, but there is still no best practice on how reward functions should be developed. We take inspiration from recent work that used meta-learning via gradient descent to construct reward functions~\cite{zou2019reward}. Because our Map() functions have smaller numbers of features, we can define a rigorous means for estimating optimal reward functions and thresholds through a simpler and faster technique, Bayesian optimization~\cite{frazier2018tutorial}. 

\subsection{Reward Shaping and Training}

Provided one or more execution context data sets, Map() and Eval() functions, and SA, our algorithm initializes hyperparameter values for the reward function and HA thresholds, then simulates autonomous missions over each execution context, and evaluates performance. We use bayesian optimization~\cite{frazier2018tutorial} to find a goal-maximizing combination of Reward, SA, and HA.

\begin{equation}
    R(S_i,A_i,S_{i+1}) = \sum_{i=0}^{n} S_{i+1}^n*w_n
\end{equation}

Reward, shown above, is the sum of each normalized feature of $S_{i+1}$ multiplied by its corresponding weight $w_i$. Weights are a feature specific hyperparameter between 0 and 1. Using these properties, we can determine that the output of $R(S_i,A_i,S_{i+1})$ will be in the range $[0,\|F_i\|]$. For this reason, $T_u$ may only fall in the same range. Similarly, $T_v$ must be an integer value between $[0,V]$. Through simulation, Bayesian optimization tunes these values until user-specified accuracy and energy requirements are met. 

Bayesian optimization is a function approximation technique that minimizes or maximizes objective functions with many parameters through creatively searching their state-space~\cite{frazier2018tutorial}. Users must provide a set of defined autonomy goals $G=\{g_1,g_2,..g_n\}$ corresponding with outputs of the Eval() function. Using G we can construct a multi-variate loss function which serves as the objective function for Bayesian optimization.
\begin{figure}
    \centering
    \begin{lstlisting}
 float[] buildReward(int nFeats, Obj[] C, int numEpochs, float[] G) {
   float[] W = initWeights(nFeats)
   int T, T* = GROUP_SIZE
   int V, V* = MAX_INT
   int bestLoss = MAX_INT
   float[] maxWeights = []
   
   for i in range(numEpochs){
     float loss = 0
     for con in C{
        float[] e = sim(con,W,T,V)
        loss += F(G, e)
     if(loss < bestLoss and goalsMet(G, e)]
        bestLoss = loss
        maxWeights, T*, V* = W, T, V
     W, T, V = Bayes([W, T, V], loss)
    }
   return maxWeights;
 }
    \end{lstlisting}
    \caption{Bayesian reward shaping pseudocode. Reward Shaping seeks to find the set of hyperparameters which minimizes loss and meets goals. 
    }
    \label{fig:reward-shape}
\end{figure}

Figure 4 shows our Bayesian reward shaping process in pseudocode. Our buildReward function accepts 4 parameters: the number of features from Map() to be weighted, the subsetset of contexts to be simulated across $\mathbb{C}_t$, the number of training epochs, and the list of autonomy goals $G$. Before training, weights and thresholds are initialized. Training consists of simulating missions using each context $c \in \mathbb{C}_t$. Loss is calculated based on autonomy goals $G$ and the final simulation evaluation $e \forall con$ using objective loss function $F(e,G)$. The additive loss for all cubes is then compared to the best prior loss. If loss is less than the best previous loss and all goals are met, the hyperparameters are saved. When the last training epoch concludes, the hyperparameters with the least additive loss that met all goals are returned, providing complete HA and reward functions.

Once reward shaping has concluded, SA is retrained with a separate subset of contexts $\mathbb{C}_r$ using a similar process. Missions are simulated across $\mathbb{C}_r$ with the new reward function and evaluated against prior training examples to select the best combination. The number of retraining periods for both SA and reward shaping are specified by user-defined hyperparameters.

%% file: OnlineLearning.tex
\section{Runtime}
\label{sect:online}

Using our goal-based development techniques for AS, users can more easily develop and train MARL algorithms for their individual problems. Output SA models may, however, be insufficiently trained to operate well in specific deployments. To help build high quality systems with limited initial training data, and allow algorithms to adapt to dynamic environments, we introduce a federated online learning system for MARL algorithms within the Fleet Computer. This system allows SA and HA models for individual swarm members to diverge to maximize global reward. Using an online learning component also introduces a scheduling and resource management problem domain for executing online learning tasks at the edge. In this section we first describe our federated learning approach to MARL for AS which uses Bayesian optimization to determine model usefulness for efficient retraining. We then present our runtime scheduling and resource management system for these online learning tasks, aided by Kubernetes.

\subsection{Bayesian optimization for system-level hyperparameters}
\label{opt}

When a swarm is deployed, each member uses the original model $SA$ for pathfinding decisions. As missions complete, swarm members' future performance may benefit from retraining using the the data they sense. It is unclear, however, whether retraining will immediately help or hurt performance, and which data should be used for maximum retraining performance. To solve these problems, we retrain a set of models $SA_s$ using different subsets of collected data and weight their performance for every system. 

\begin{figure}[t]
 \centering
  \includegraphics[scale=0.90]{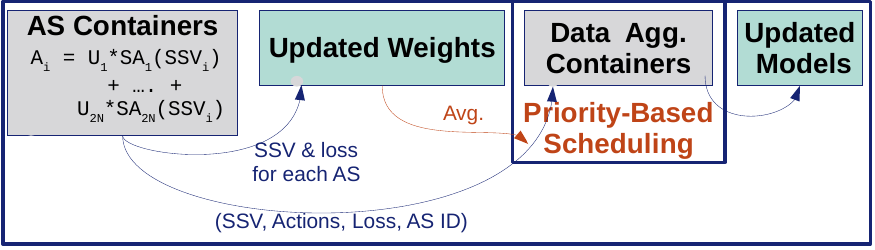}
 \caption{Online learning provides coefficients for priority-based scheduling to update MARL models.}
  \label{fig:runtime}
\end{figure}

Our solution to this problem is depicted in \MYfigureref{fig:runtime}. After a deployment's first mission, each swarm member has sensed and stored new data that can be used by aggregation containers for retraining. The sensed data can be partitioned into an infinite number of subsets to retrain $SA \rightarrow SA_s$.
Using intuition from the federated learning community ~\cite{konevcny2016federated}, we define the upper bound of $SA_s$ as the set of models retrained using all combinations the powerset of $N$, $\mathbb{P}(N)$. This gives us latitude in selecting optimal models, but provides resource challenges for large swarms. The Fleet Computer provides developers the option to provide heuristics to prune aggregators from $SA_s$, and can use usefulness to prune $SA_s$ online. It is unclear a priori how each model in $SA_s$ will affect each swarm member, so training effects must be determined online. Therefore, we track a series of performance-based system level hyper-parameters.

We track model weights $\{U_w^1..U_w^i\}$ for models $\{SA_1..SA_i\} \in SA_s$. Model weights describe the weighted percentage ($0 \leq U_w^x \leq 1$, $\sum_{n=1}^{i} U_w^n = 1$) that decisions from that model are incorporated into a pathfinding decisions. The goal of our approach is to determine the values for these hyperparameters which yield us the best mission performance. Per swarm member, we define these hyperparameters as $x$.

\begin{equation}
x = \{U_w^1, U_w^2, ..U_w^i\}
\end{equation}

$SA_s(x)$ describes the Fleet Computer mission whose pathfinding decisions for a single agent are determined by $SA_s$ weighted by x. The Fleet Computer's goal is to determine a set of hyperparameters $x^*$ that best accomplishes autonomy goals. More specifically, we want to optimize the following objective function.

\begin{equation}
    x^* = x \: s.t. \: argmax(Eval(SA_s(x)) | SA_s(x) > G)
\end{equation}

We use $Eval(SA_s(x))$ representing the evaluation of a full mission informed by $SA_s(x)$. Our goal is to find x*, the set of weights that maximize the evaluation, while assuring that our autonomy goals $G$ are met.

Bayesian optimization~\cite{mockus1991bayesian,frazier2018tutorial} is a powerful technique often used in hyperparameter tuning to improve machine learning
algorithms performance. Bayesian optimization works well for optimization problems where objective functions are derivative free, expensive to evaluate, noisy, lack structures that are easy to optimize, and have small parameter sets~\cite{frazier2018tutorial}. Reinforcement learning approaches often fit all of these criteria, ours included. To optimize our hyperparameters, we use expected improvement to quickly minimize our objective function given specified goals. Expected improvement approximates the global maximum of our objective function by sampling hyperparameters based on the posterior distribution of prior sample points and loss. Expected improvement determines hyperparameter samples with a balance between the highest probability for improvement and the largest magnitude of that improvement. We model this as a Gaussian process.

The above approach frames Bayesian optimization in the scope of a single agent's parameters, but we can use this approach to optimize hyperparameters for multiple agents. Using the same local-optimization approach, we can determine a near-optimal $x$ to maximize global goals. 
Each swarm member maintains its own set of hyperparameters $x_i$ which is updated asynchronously after each mission. The collective set of hyperparameters provides us a simple metric for determining how 'useful' each model in $SA_s$ is to overall pathfinding decisions. Models with higher average $U_w^i$ provide more insight into pathfinding decisions, and therefore should be updated more regularly and provided more retraining resources.

\subsection{Scheduling and power management}
\label{sect:implementation}

The Fleet Computer consumes edge resources to support movement, actuation, pathfinding, data storage, and online learning. Within this set of activities it is important to balance the resource needs of critical real-time latency-sensitive processes, like movement control, with compute intensive tasks like retraining routines that offer significant long-term gain. It is also important for edge resources to be responsive to workload changes, powering down edge devices in low load periods to maximize system liveness and mission length, and scaling out in peak loads to support effective decision-making. Here we describe our solutions to both of these challenges: Section~\ref{ssect:scheduling} presents our priority-based container scheduling solution for latency-sensitive and compute-intensive tasks, which factors in utility of each compute-intensive training task to overall mission objectives, and Section~\ref{ssect:scaling} describes our cluster autoscaling mechanism to adjust to workload changes.

\subsubsection{Priority-based Scheduling}
\label{ssect:scheduling}

The Fleet Computer deploys all of its core components in containers~\cite{merkel2014docker} to maintain hardware independence and support scale-out. Figure~\ref{fig:arch} shows the different container types that encapsulate pre-built inputs like MARL algorithms, retraining routines, and feature extractors along  with core Fleet Computer platform elements.

Container scheduling across clusters is well understood~\cite{casalicchio2019container}. We use Kubernetes~\cite{sayfan2017mastering} for automated bin-backing of containers across clusters, allocating resources for all Fleet Computer components defined as Docker containers. We use priority-based scheduling to ensure that latency-sensitive real-time tasks such as  UAV flight control are assured to be executable within their latency window; we then use additional edge compute to schedule model retraining tasks based on newly-received data from swarm members of the in-progress mission.

For model retraining tasks, we augment the resource allocation algorithm used by Kubernetes to optimise \textit{placement} and \textit{task selection} for Fleet Computer operations. Placement decisions are impacted by data locality, where training data for SA and HA models is typically fragmented across multiple systems within the Fleet Computer. Our resource allocation algorithm guarantees that containers will be scheduled on an edge node that either (1) has at least some of the data required for model retraining, or (2) is within a user-defined edge hub with expanded compute. This provides Kubernetes with sufficient flexibility in scheduling, but guarantees that data transfer times remain relatively low and allows for the potential of partial or complete data locality at the training site.

Task selection is impacted by the likely utility of a retraining task: when edge compute resources cannot support all retraining tasks, we choose those with the highest average utility. The Fleet Computer uses Kubernetes' priority scheme for scheduling training procedures using the aggregate usefulness of each model provided by AS containers as shown in \MYfigureref{fig:runtime}. Model usefulness is simply the floating point value $U_i=[0,1]$ assigned by Bayesian optimization, as explained in Section~\ref{opt}. If model usefulness is calculated by multiple swarm members, usefulness is weighted evenly among them. Scheduler priority for model $i$ is then:

\begin{equation}
    P_i = \round{10 * U_i}*100,000,000
\end{equation}

$P_i$ then becomes a priority value in range $[0..900,000,000]$ at intervals of $100,000,000$, allowing for 10 possible priority values. This range maps into the entire range of priority levels available in Kubernetes, which is specified by integers between 0 and 1 billion, while providing a coarse granularity that clearly differentiates retraining routines of different utilities and allows us to reserve the priority level 10 for sensor containers. 

We also use usefulness to assign compute resources. Kubernetes allows users to provide minimum and maximum resource usage constraints when pods are instantiated. We use the same priority numbers $[0,9]$ to assign relative CPU and RAM maximums to pods. All available RAM and CPU units are portioned among pods based on their priority levels.

\begin{equation}
    CPU_i = \frac{CPU_t}{\sum_{j=0}^{s}P_j} * P_i
\end{equation}

Shown above, all CPU cores available across the system $CPU_t$ for retraining are split evenly among pods based on priority. This same process is used to allocate memory. Pods with a priority of 0 are not scheduled. This priority mechanism allows us to provide more resources to retrain models that agents consider useful, and avoid retraining models that provide little to the system. 

\subsubsection{Cluster Autoscaling}
\label{ssect:scaling}

Because Fleet Computers may include many nodes, and edge devices in particular are likely to be provisioned for peak loads rather than average load, it is beneficial to regularly scale the cluster up and down in response to workloads.

The Fleet Computer includes a custom Kubernetes autoscaler which drains compute tasks from superfluous nodes and powers them down, saving on edge power when loads are low, and re-powers decommissioned nodes using Wake-On-LAN~\cite{mishra2006wake} when the system detects that more compute will soon be required.

Powering down nodes in this way must be sensitive to cluster storage implications. Each edge node stores different fragments of data, so we must ensure that no data becomes unavailable. For this purpose we use the Hadoop Distributed File System (HDFS)~\cite{shvachko2010hadoop} for cluster data management, configured to replicate all data twice across the edge node cluster. When data is `lost' from a decommissioned node it will therefore still be available at one other node in the cluster; after each decommission we simply wait for data to be re-replicated by HDFS before powering down any further nodes, guaranteeing data availability as the active node population changes.

%% file: Implementation.tex
\section{Applications}
\label{sect:impl}

We use two very different application types to help evaluate the Fleet Computer: one using a swarm of crop scouting UAVs, and one using a swarm of taxi tracking cameras.

Our crop scouting swarm builds on prior work~\cite{zhang2020} where UAVs navigate a corn field and capture images which are used to construct a yield map as outlined in Section 2. Farmers use yield maps to inform crop management strategies like fungicide, pesticide, and herbicide application. Large crop fields may cover hundreds of acres, requiring days of swarm coverage and human labor to gather a complete yield map. Prior work used reinforcement learning to creatively sample fields, covering a fraction of the field autonomously and predicting the rest. This approach creates usable yield maps in a fraction of the time at lower cost, but requires considerable developer effort. Actions in the crop scouting application are UAV direction movements, in a field which is divided into a grid. In this application we used the same dataset of corn images as prior work, consisting of 684 UAV sensed 4608x3456 images of a corn field in London, Ohio. We input the same Q-learning based model and feature extraction from prior work into the fleet spec as $SA$ and $Map()$, and modified the extrapolation function from prior work which generated crop yield maps to produce an $Eval()$, which provides goal information like edge and UAV energy along with overall map accuracy. We used the Fleet Computer's reward shaping mechanism to build reward and distance functions, defining state-groups as unique 3x3 regions of states.

Our taxi tracking swarm also builds on prior work in video analytics~\cite{jain2020spatula}. 
Spatula is a cross-camera video analytics framework: for a specific target (a person, taxi, etc) in one video stream, it returns all frames across all cameras which contain that target. Spatula avoids searching all cameras by searching only cameras that are highly spatially and temporally correlated. Spatula builds these correlation matrices offline using prior execution data. Online, cameras are only searched if their spatial and temporal correlation scores are higher than user-provided thresholds. This approach performs considerably better than searching all cameras and frames, but runs the risk of staleness as movement patterns change over time, and also requires users to manually determine threshold values. To implement this application we used the Porto Taxi Service Trajectory dataset~\cite{PORTO}, also used to test Spatula. Similar to the original implementation, we generated spatial and temporal correlations using 130 virtual cameras pinned in an evenly spaced grid within Porto, Portugal. Correlation matrices were provided to the Fleet Spec as the State-to-Action model, and trajectories from the Porto data set were provided in place of $Map()$. We built a custom $Eval()$ function which reports overall accuracy and the total number of frames checked for targets throughout execution, and we generated an HA function to determine optimal temporal and spatial thresholds. We defined state-groups as 50\% overlapped 1-minute sets of video frames from each camera.

Online learning for both systems consists of retraining models and rebuilding reward and distance functions using recently received execution data. For crop scouting, swarm sizes were small, so we dispatched $\mathbb{P}(N)$ aggregators for retraining after every crop-scouting mission (i.e when a swarm generates a final yield map). For Spatula, swarm sizes are large, so we dispatched $N$ aggregators after every simulated day to rebuild correlation matrices and relearned thresholds.

%% file: Results.tex
\section{Evaluation}
\label{sect:results}

We implemented swarms for both applications described in Section~\ref{sect:impl} 
using multiple competing approaches. In this section, we describe 
our experimental testbed and then evaluate the efficacy and performance of Fleet Computer swarms.


\begin{figure*}[t]
 \centering
  \includegraphics[scale=0.9]{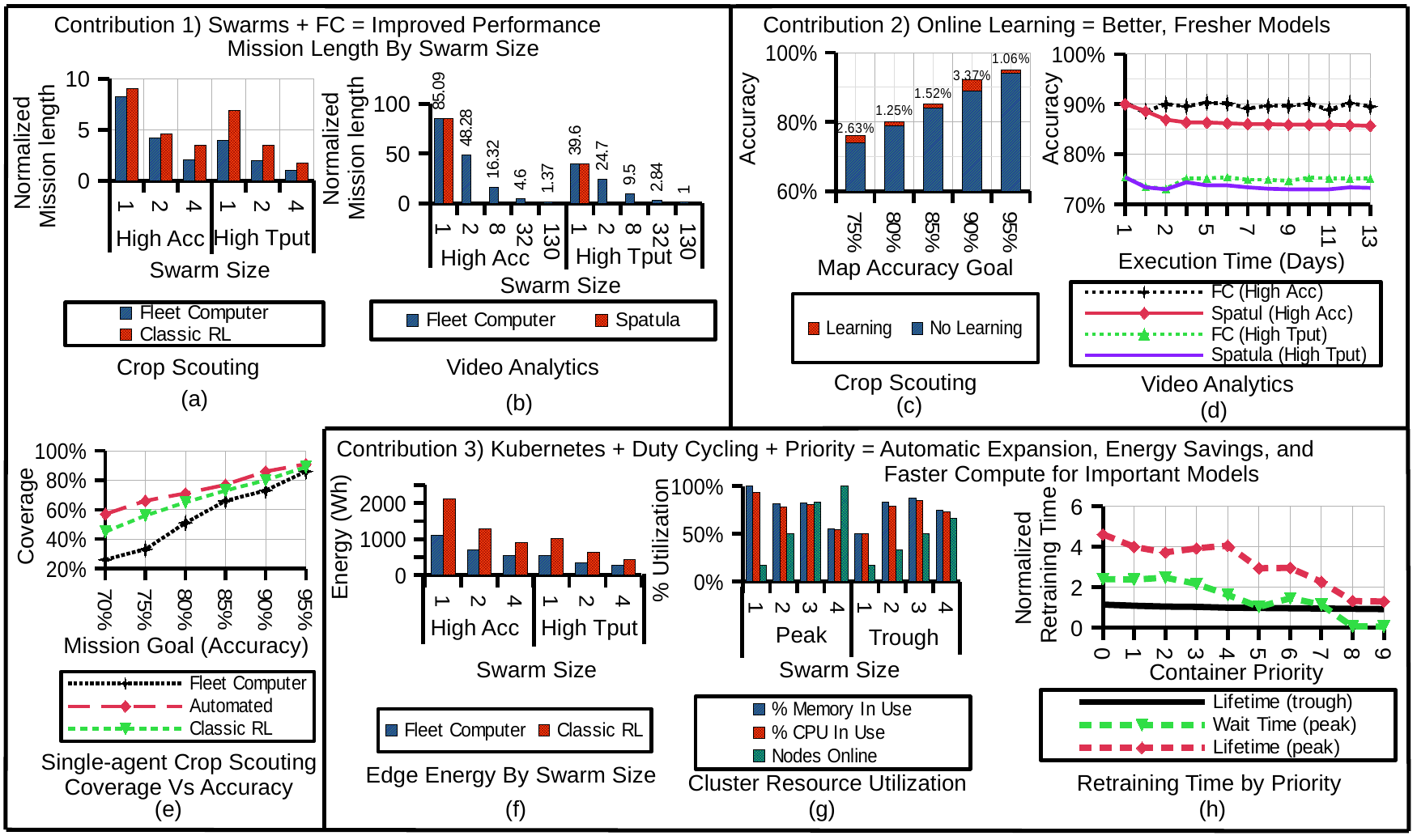}
 \caption{Experimental results: a,b,e) The Fleet Computer's programming model and swarm capabilities improve performance considerably, c-d) Online learning keeps models fresh and helps adapt to new execution contexts, f-h) The Fleet Computer's system management features save energy and improve performance by duty cycling hardware, expanding compute efficiently, and prioritizing high-value model updates.}
  \label{fig:autonomy-types}
\end{figure*}

\subsection{Architecture}

The Fleet Computer is designed for edge deployments.  
Our canonical prototype uses 6 total machines, 
comprising 5 consumer laptops as gateways and a 
server as an edge hub. Consumer laptops include 
3 HP 250 G6 laptops with i5 CPUs and 8GB RAM and two Lenovo Thinkpad T470s with i7 CPUs and 8GB RAM. 
The server includes an i9 CPU, 32GB RAM, and an Nvidia RTX 3080 GPU. Each machine runs Ubuntu 20.04 Linux. All systems in the Fleet Computer are connected by Ethernet to a 10 Gbps Netgear router.

The Fleet Computer uses one of the Lenovo Thinkpads as a master node to control Kubernetes, Docker, and HDFS. The master node also runs a custom Kubernetes governor, a collection of daemons that create the Kubernetes cluster, start autonomous missions, automatically allocate and prioritize retraining containers, autoscale the cluster, and duty-cycle other machines. When retraining occurs, containers are scheduled by the governor as pods in the Kubernetes cluster. Each pod is scheduled based on its aggregate usefulness as determined by all swarm members as discussed in Section~\ref{sect:impl}.

For our crop scouting application, DJI Mavics provide the base UAV characteristics for emulation.  
Each UAV in a swarm is assigned an HP or Lenovo laptop as a gateway for communication with the Fleet Computer.
The UAV's control software runs in a container on the gateway.  
To gather crop scouting results, we simulated UAV by replaying data captured from previous missions. All UAV control software executed as it would in the field, but data was provided by software. This allowed us to execute all UAV commands without flying, but receive appropriate execution data based on validated energy and latency models~\cite{boubin2019managing}. 

\subsection{Results}

We evaluate the Fleet Computer's comparative performance against state-of-the-art swarm control, the effectiveness of our runtime scheduling approach in saving energy on edge devices, and the effectiveness of our online learning approach at autonomously selecting effective data aggregations for retraining on our resource-constrained edge devices.

Figure~\ref{fig:autonomy-types} (a-b) shows the Fleet Computer's performance on our crop scouting and video analytics workloads \textit{without} additional online learning. For crop scouting, we compare against the state of the art in prior work~\cite{zhang2020} (\textit{Classic RL}) which utilized Q-learning to map crop fields. We test the Fleet Computer against classic RL using 3 swarm sizes (1, 2, and 4 UAVs), and two autonomy goals: one seeking high-accuracy (>90\%) maps; and the other prioritizing throughput, accepting 70\% accurate maps in exchange for fast sampling. Both approaches started with the same Q-learning State-to-Action model, but the Fleet Computer generates its own History-to-Action model and reward function from experience.

For the both accuracy conditions, the Fleet Computer decreased mission length by 10-75\% compared to Classic RL. Using a swarm of 4 UAVs, compared to a single UAV as used in prior work, we observe a mission length decrease of 3.9-4.4X. This decrease is in part due to having more UAVs, but represents more than 4x the performance gain due to the Fleet Computer's intelligent mission learning and redesign.


Figure~\ref{fig:autonomy-types}(e) shows why Fleet Computer missions complete more quickly -- compared against both a Classic RL and current-industry-practice automated search which scouts waypoints linearly, row by row until the coverage goal is reached~\cite{hu2020hivemind,barrientos2011aerial}. The Fleet Computer consistently outperforms both prior approaches, most notably at lower accuracy goals, by sampling an  average of just 26\% of a crop field to generate a 70\% accurate yield map, improving over Classic RL and automated by 1.73X and 2.2X respectively. These improvements are gained because the Fleet Computer learns from experience which areas of a field are likely to be similar to adjacent regions and so do not need detailed data capture.

Figure~\ref{fig:autonomy-types}(b) shows the Fleet Computer's performance against Spatula. Spatula uses a shared correlation matrix and a single controller to implement cross-camera analysis, and so is only shown in the `1' category on Figure~\ref{fig:autonomy-types}(b). The Fleet Computer enables us to easily model cross-camera analytics as a distributed swarm of cameras, where each agent is provided compute resources to respond to global queries across one or more cameras. Using this approach, the Fleet Computer can highly parallelize query response and so offers very large potential performance gains without complex programming. We used the Fleet Computer's History-to-Action model to automatically set Spatula's spatial and temporal thresholds, and tested performance by swarm size and for high throughput and high accuracy goals. Our experiments examine performance gains at increasing numbers of swarm agents applied to the same problem, which is easy to configure through the Fleet Computer; at the highest number of 130 swarm agents we found that the Fleet Computer can parallelize Spatula's workload highly effectively, processing queries 39X and 62X faster than the single Spatula controller.

When we introduce \textit{online learning} to these scenarios we see further improvements for both applications. Figures~\ref{fig:autonomy-types}(c) and (d) show how crop-scouting can improve over time, and how regularly updating HA models can maintain Spatula's performance as movement patterns evolve.

Figure~\ref{fig:autonomy-types}(c) shows how online learning improves crop scouting accuracy. In this experiment we run 10 successive real-time swarm missions across simulated crop fields, allowing online learning to continually improve models across a 4 UAV swarm for accuracy goals between 75\% and 95\%. By the 10th mission we see that every condition moves closer to its target accuracy level by between 1\% and 3\%; if we measure this as decrease in relative model error from the target accuracy this equates to between 5\% and 28\% improvement.
These savings improve accuracy goals with no increase in resource requirements or mission lengths.

Figure~\ref{fig:autonomy-types}(e) shows that Spatula's performance with retraining remains consistent with autonomy goals if retrained, but degrades quickly without retraining. We trained Spatula on one day of Porto Taxi Data using all 130 cameras and 448 Taxis, then examined its performance compared to the Fleet Computer on each following day of a two-week period after its initial training. For both high accuracy and high throughput goals, the Fleet Computer remains consistent (within 1\%) with accuracy goals. Spatula, in both cases, consistently under-performs accuracy goals after 2 and 4 days for high accuracy and throughput respectively. At the end of the two week period, the Fleet Computer outperforms Spatula by 5\% and 3\% for high accuracy and throughput respectively by maintaining fresh models.

We next examine the Fleet Computer's resource management approach and its effect on edge site energy usage. Figures~\ref{fig:autonomy-types}(f-h) show how the Fleet Computer's edge-focused Kubernetes runtime improves overall edge performance on our crop scouting benchmark. Using the prototype Fleet Computer, we ran 10 real-time crop scouting missions for each swarm setting using UAV-collected data with modeled UAV movement, timings, and data-transfer provided by prior work~\cite{boubin2019managing}. Energy was calculated using an AC watt meter. We tested 1, 2, and 4 drone swarms with both high accuracy and high throughput goals. 
Figure~\ref{fig:autonomy-types}(f) shows the Fleet Computer's performance against Classic RL in terms of energy consumption. The Fleet Computer conserves energy in two different ways: mission lengths are shorter overall, and use of edge sites during a missing is cheaper due to automated power-down of resources in non-peak load periods. Compared to Classic RL, similar sized Fleet Computer missions consume 1.58X-2X less power. A swarm of multiple UAVs is also more energy efficient per-device than fewer UAVs; compared to Classic RL using a single UAV as in prior work, a swarm of four fleet-computer-controlled UAVs uses 3.7X-3.9X less energy depending on swarm size and goals.

Figure~\ref{fig:autonomy-types}(g) shows how the Fleet Computer manages resources across extremes during a swarm mission. For a single UAV, one node of the Fleet Computer cluster is enough to handle all resource needs. As the swarm grows, more nodes must be provisioned. As nodes increase from 2 to 4, peak and trough allocation change. For instance, a 4-UAV swarm can operate at troughs using just 4 nodes, but requires all 6 to handle peak loads. The Fleet Computer's energy savings shown in Figure~\ref{fig:autonomy-types}(f) are a direct result of its ability to spread resources evenly across the cluster and shut down unnecessary nodes until they are needed. 

Finally, we examine the Fleet Computer's usefulness-aware model retraining; this offers better use of finite edge resources by selecting which training tasks are likely to yield the highest benefit to an ongoing mission. Figure~\ref{fig:autonomy-types}(h) shows how our usefulness metric affects model retraining times compared to average retraining times. When the system is not under load, Kubernetes is easily able to distribute containers across it. If the system is correctly provisioned for the edge, however, it may experience peak loads where containers must contend for resources. We evaluated retraining times for 4-UAV on our crop-scouting benchmark. We found that wait-times for high-priority containers (8-9 on the x-axis) were insignificant even at peak loads, but could be up to 2.4X normal retraining time for very low (0-2) priority containers. Similarly, high-priority containers experienced only modest (1.3X) lifetime increases even when the system was highly pressured. This was at the expense of lower priority containers, which experienced lifetimes up to 4.6X longer than usual. This behavior allows the system to take resources from models with low utility to the system and give them to high utility models. 

%% file: Discussion.tex
\subsection{Limitations and Future Work}

The Fleet Computer uses Bayesian optimization for 
reward shaping, online learning and scheduling.
While Bayesian optimization has admirable attributes, 
like fast convergence, other popular techniques, such as 
gradient descent, could be applied as well.  We have not 
explored the tradeoffs in convergence time and learning efficacy.  
HA models leverage spatial properties that most MARL algorithms do not, 
but our approach is relatively simple.  A more complex model could 
further reduce unneeded actions, but may also increase training and inference times. 
In particular, FlexDNN provides a rigorous analysis of 
similar early-exit models in DNN frameworks~\cite{fang2020flexdnn}.
We evaluated two applications that are analogous to Q-learning. 
Future work could explore deep Q-networks increasingly used in autonomous systems. 
Finally, while we used real hardware and workloads in our evaluation, 
we simulated data capture from edge devices and did not explore 
swarms bottlenecked by data ingestion.

%% file: relatedwork.tex
\section{Related Work}
\label{sect:rw}

Much recent work has dealt with tackling the concerns of real-world autonomous systems using strong theoretical foundations. Lin et al~\cite{lin2020collaborative} explores a federated meta-learning approach to train models with small datasets in an edge setting. Singh et. al~\cite{singh2019end} demonstrates a novel reward-training mechanism for reinforcement learning to eliminate the need for reward shaping. Kilinc and Montana~\cite{kilinc2018multi} constructs a framework for sharing data among agents in execution contexts that are noisy and non-stationary using intrinsic reward and temporal locality. Other recent work~\cite{zhang2018fully,zhang2018networked,qu2020scalable} on networked agents provide considerable insight into the behaviors of real-world cooperative MARL systems with limited communication capabilities. Porter et. al~\cite{porter2016rex} presents a novel development platform for creating software that autonomously assembles itself and discovers optimal execution policies online without the need for expert model building and reward shaping.

Much related work deals specifically with autonomous aerial systems. Boubin et. al~\cite{boubin2019managing} demonstrates that naive hardware and algorithm selection for fully autonomous aerial systems can have serious performance consequences. Cui et. al~\cite{cui2019multi} implements MARL for allocating networking resources across a network of UAV base-stations. In agriculture Zhang et. al~\cite{zhang2020}, Yang et. al~\cite{yang2020adaptive}, and FarmBeats~\cite{vasisht2017farmbeats} provide new techniques for automated and autonomous UAV crop scouting. 

%% file: conclusion.tex
\section{Conclusion}
\label{sect:conclusion}
Swarms of autonomous systems powered by resources at the edge can provide insight and actuation that will revolutionize industries like agriculture, construction, transportation, and video analytics. 
We present the Fleet Computer, an end-to-end platform 
for building, deploying, and executing swarms.  
To use the Fleet Computer, developers implement 
{\em Map()} and {\em Eval()} functions and specify
mission configurations.  The Fleet Computer compiles and 
deploys swarms using a multi-agent reinforcement learning 
framework.  At runtime, the Fleet Computer manages data aggregation
between swarm members, linking online learning outcomes 
to the efficient management of edge resources.
Our evaluation showed that the Fleet Computer can produce
effective and efficient swarms, suggesting this tool chain
could make swarms accessible to everyday developers.

